\documentclass[twocolumn,showpacs,aps,prl,superscriptaddress]{revtex4b5}

\usepackage{graphicx}
\usepackage{dcolumn}
\usepackage{amsmath}
\usepackage{epsfig}

\input pubboard/babarsym
\def\runIlumi   {\ensuremath{20.7 \invfb}}
\def\Btopsikst     {\ensuremath{ B \to \jpsi \Kstar}}
\def\Btopsikstz    {\ensuremath{\Bz \to \jpsi \Kstarz}}
\def\Btopsikstp    {\ensuremath{\Bu \to \jpsi \Kstarp}}

\def\ksttokspiz {\ensuremath{\Kstarz \to \KS \piz}}

\def\az    {\ensuremath{A_{0}}}
\def\ap    {\ensuremath{A_{\parallel}}}
\def\at    {\ensuremath{A_{\perp}}}
\def\rt    {\ensuremath{R_{\perp}}}

\def\azd   {\ensuremath{|\az|^{2}}}
\def\apd   {\ensuremath{|\ap|^{2}}}
\def\atd   {\ensuremath{|\at|^{2}}}

\def\phip  {\ensuremath{\phi_{\parallel}}}
\def\phit  {\ensuremath{\phi_{\perp}}}
\def\thetakstar {\ensuremath{\theta_{\Kstar}}}
\def\phitr      {\ensuremath{\phi_{tr}}}
\def\thetatr    {\ensuremath{\theta_{tr}}}
\def\cthetatr   {\ensuremath{\cos{\thetatr}}}

\def\cthetakstar{\ensuremath{\cos{\thetakstar}}}

\def\sphitr     {\ensuremath{\sin{\phitr}}}
\def\pipt       {\ensuremath{\text{Im}{(\ap^{*}\at)}}}
\def\przp       {\ensuremath{\text{Re}{(\az^{*}\ap)}}}
\def\pizt       {\ensuremath{\text{Im}{(\az^{*}\at)}}}
\def\avEps    {\ensuremath{ \langle\epsilon\rangle }}

\newcommand{\cq}[1]{\cos^{2}{#1}}
\newcommand{\sq}[1]{\sin^{2}{#1}}
\newcommand{\gfrac}[2]{\displaystyle\frac{#1}{#2}}
\newcommand{\dd}{\text{d}}

\def\gobs       {\ensuremath{g_{obs}}}

\newcommand{\BABARPubYear}    {01}
\newcommand{\BABARPubNumber}  {05}

\newcommand{\SLACPubNumber} {8898}

\def\figurebox#1#2#3{%
    \def\arg{#3}%
    \ifx\arg\empty
    {\hfill\vbox{\hsize#2\hrule\hbox to #2{\vrule\hfill\vbox to #1{\hsize#2\vfill}\vrule}\hrule}\hfill}%
    \else
    {\hfill\epsfbox{#3}\hfill}%
    \fi}

\long\def\inst#1{\par\nobreak\kern 4pt\nobreak
    {\it #1}\par\vskip 10pt plus 3pt minus 3pt}

\begin{document}


\begin{flushleft}
\babar-PUB-\BABARPubYear/\BABARPubNumber\\
SLAC-PUB-\SLACPubNumber\\
\end{flushleft}

\title{
\vskip 10mm
{\large \bf Measurement of the $\Btopsikst(892)$ decay amplitudes}
\begin{center} 
\vskip 10mm
The \babar\ Collaboration
\end{center}
}

%
\author{B.~Aubert}
\author{D.~Boutigny}
\author{J.-M.~Gaillard}
\author{A.~Hicheur}
\author{Y.~Karyotakis}
\author{J.~P.~Lees}
\author{P.~Robbe}
\author{V.~Tisserand}
\affiliation{Laboratoire de Physique des Particules, F-74941 Annecy-le-Vieux, France }
\author{A.~Palano}
\affiliation{Universit\`a di Bari, Dipartimento di Fisica and INFN, I-70126 Bari, Italy }
\author{G.~P.~Chen}
\author{J.~C.~Chen}
\author{N.~D.~Qi}
\author{G.~Rong}
\author{P.~Wang}
\author{Y.~S.~Zhu}
\affiliation{Institute of High Energy Physics, Beijing 100039, China }
% ...
\collaboration{The \babar\ Collaboration}
\noaffiliation

%

\date{\today}

\begin{abstract}
We present a measurement of the decay amplitudes in $\Btopsikst(892)$
channels using \runIlumi\ of data collected at the \FourS\ resonance with
the \babar\ detector at \pep2.
We measure a $P$-wave fraction $\rt = (16.0\pm 3.2 \pm 1.4)\%$
and a longitudinal polarization fraction $(59.7\pm 2.8 \pm 2.4)\%$.
The measurement of a relative phase that is neither 0 nor $\pi$, $\phip = 2.50\pm 0.20 \pm 0.08$ radians,
favors a departure from the factorization hypothesis.
Although the decay $\B\to \jpsi K\pi$ proceeds mainly via $\Kstar(892)$,
there is also evidence for $\Kstar_2(1430)$ and $K\pi$ $S$-wave contributions.
\end{abstract}

\pacs{12.15.Hh, 11.30.Er, 13.25.Hw}

\maketitle

The decay \Btopsikstz\ with \ksttokspiz\ allows a measurement of the
\CP\ violation parameter \stwob that is theoretically as clean as for
\bpsiks\ \cite{ref:bigi}.
However, due to the presence of even $(L=0,2)$ and odd $(L=1)$ orbital angular momenta
in the \jpsi \Kstar system, there can be \CP-even and \CP-odd contributions
to the decay rate.
If the information contained in the decay angles is ignored, 
the measured time-dependent \CP\ asymmetry is reduced
by the dilution factor $D_\perp = 1 - 2\rt$, where \rt\ is the fraction of $P$-wave.
If the angular information is used, the \CP\ components can be separated \cite{ref:dunietz}.

The angular analysis also provides a test of the factorization hypothesis, the validity of
which is in question for color-suppressed modes \cite{ref:yeh,ref:rosner2000kx}.
In this scheme, the weak decay is described by a product of $\jpsi$ and \B \to \Kstar
hadronic currents, and final state interactions are neglected. If factorization holds,
the decay amplitudes should have relative phase $0$ or $\pi$.

The decay $\Btopsikst(892)$ is described by three amplitudes. In the transversity basis
\cite{ref:dunietz,ref:dighe} used by CLEO \cite{ref:cleo} and CDF \cite{ref:cdf},
the amplitudes \ap, \az\ and \at\ have \CP\  eigenvalues $+1, +1$ and $-1$, respectively.
\az\ corresponds to longitudinal polarization, and
\ap\ and \at\  respectively to
parallel and perpendicular transverse polarizations, of the vector mesons;
\rt\ is defined as $\atd$.
For a $\Delta I = 0$ transition, all $\Kstar\to K\pi$ channels
involve the same amplitudes, and so the data for different decay modes can
be combined.

The transversity frame is defined in the \jpsi\ rest frame.
The \Kstar\ direction defines the negative $x$ axis.
The $K\pi$ decay plane defines the $(x,y)$ plane, with $y$ such that $p_y(K) > 0$. The $z$ axis is
the normal to this plane, and the coordinate system is right-handed. 
The transversity angles \thetatr\ and \phitr\ are
defined as the polar and azimuthal angles of 
the positive lepton from \jpsi decay; \thetakstar\ is the 
\Kstar\ helicity angle defined as the angle
between the $K$ direction and the direction opposite the \jpsi\ in the \Kstar\ rest frame.
The normalized 
angular distribution $g(\cthetatr, \cthetakstar, \phitr)$ is
\begin{eqnarray}
\label{eqn:distrib}
g & = & \gfrac{1}{\Gamma}\gfrac{\dd^{3}\Gamma}{\dd\cthetatr\;\,\dd\cthetakstar\;\,\dd\phitr} \\
                                   & = & f_1\cdot\azd  + f_2\cdot\apd  + f_3\cdot\atd    \nonumber \\
                                   & + & f_4\cdot\pipt + f_5\cdot\przp + f_6\cdot\pizt   \nonumber
\end{eqnarray}
with
\begin{eqnarray*}
f_1 & = & ~ ~ \ \, 9/(32\pi)\cdot\,  2\cq{\thetakstar}(1-\sq{\thetatr}\cq{\phitr}), \\
f_2 & = & ~ ~ \ \, 9/(32\pi)\cdot\, \sq{\thetakstar}(1-\sq{\thetatr}\sq{\phitr}), \\
f_3 & = & ~ ~ \ \, 9/(32\pi)\cdot\, \sq{\thetakstar}\sq{\thetatr}, \\
f_4 & = & ~ ~ \ \, 9/(32\pi)\cdot\, \sq{\thetakstar}\sin{2\thetatr}\sphitr\cdot\zeta, \\
f_5 & = & - \ \, 9/(32\pi)\cdot\,  1/\sqrt{2}\cdot\sin{2\thetakstar}\sq{\thetatr}\sin{2\phitr}, \\
f_6 & = & ~ ~ \ \, 9/(32\pi)\cdot\,  1/\sqrt{2}\cdot\sin{2\thetakstar}\sin{2\thetatr}\cos{\phitr}\cdot\zeta.
\end{eqnarray*}
When the final state is not a \CP\ eigenstate, $\zeta$ is $+1$ for \Bu\ and \Bz , and $-1$ for
\Bub\ and \Bzb  .
For the \CP\ mode $\KS\piz$, $\zeta(\Bz) = -\zeta(\Bzb) = 1/(1+x_d^2)$,
where $x_d = \Delta m_{B_d}/\Gamma_{B_d} \sim 0.73$;
however, since flavor is not determined in the present analysis,
$\zeta$ averages to zero for this mode. We define the relative phases of the amplitudes as 
$\phit=\arg(\at/\az)$ and $\phip=\arg(\ap/\az)$.

In this letter, we present a measurement of the decay amplitudes in the decays
\Btopsikstz\ and \Btopsikstp , where the \Kstarz\ and \Kstarp\ are reconstructed in the modes
$\KS\piz$, $\Kp\pim$ and $\KS\pip$, $\Kp\piz$, 
respectively \cite{foot}; 
Only \jpsi\ decays to \epem\ and \mumu\ are considered.
The data sample corresponds to \runIlumi\ collected at the \FourS\ in
1999--2000 with the \babar\ detector at the \pep2\ \abf, and contains
$\sim 22.7\times10^6$ $B$ meson pairs.

The \babar\ detector is described elsewhere \cite{ref:babardet}.
Charged particle track parameters are obtained from measurements in a 5-layer double-sided
silicon vertex tracker and a 40-layer drift chamber located in a 1.5-T magnetic field;
both devices provide \dedx\ information.
Additional charged particle identification (PID) information is obtained from a
detector of internally reflected Cherenkov light (DIRC) consisting of
quartz bars that carry the light to a volume 
filled with water, and equipped with 10752 photomultiplier tubes.
Electromagnetic showers are measured in a calorimeter (EMC) consisting of 6580 CsI(Tl)
crystals.
An instrumented flux return (IFR), containing multiple layers of resistive plate chambers,
provides $\mu$ identification.

Electrons are identified by requiring that shower shape and
energy deposition in the EMC be compatible with
those expected for an electron of the measured momentum;
\dedx\ measurements must also be compatible with the electron hypothesis. 
Muon candidates must
penetrate at least two interaction lengths in the detector,
and generate a small number of hits per layer in the IFR.
If a muon candidate traverses the EMC,
its energy deposition must be consistent with that of a minimum ionizing particle.
Kaon candidates must survive a pion veto based on DIRC and \dedx\ information.

\begin{figure}[t]
 \begin{center}
 \includegraphics[width=0.93\linewidth]{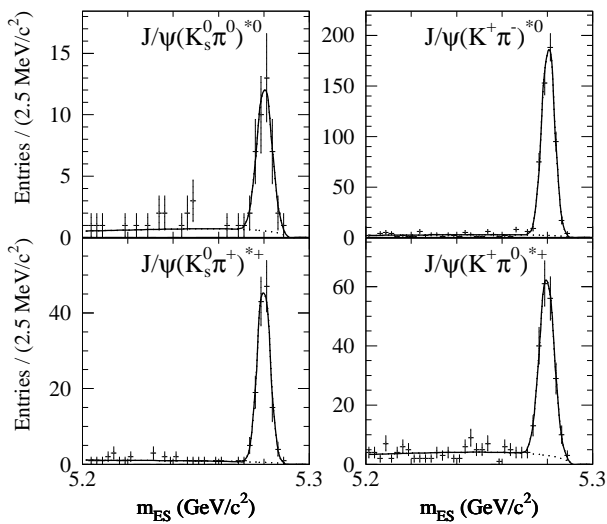}
 \caption{Beam-energy substituted mass spectra for the four $K\pi$ modes.
The curves are from fits using the $G(\mes)$ and $F(\mes)$ functions
described in the text.}
 \label{fig:bmasses}
 \end{center}
\end{figure}

Charged tracks are required to be in regions of polar angle for which the
PID efficiency is well-measured.
For electrons, muons and kaons the acceptable ranges are 0.41 to 2.41, 0.3 to 2.7 and
0.45 to 2.5\rad, respectively.
 \jpsi\ candidates  consist of a pair of identified leptons that form a good vertex.
The lepton pair invariant mass must be between 3.06 and 3.14\gevcc\ 
for muons and 2.95 and 3.14\gevcc\ for electrons.
This corresponds to a $\pm3\sigma$ interval for muons, and  accounts for
the radiative tail due to bremsstrahlung for electrons.
\KS\ candidates consist of vertexed pairs of oppositely-charged tracks
with invariant mass between 489 and 507\mevcc.
In the plane perpendicular to the beam line, the \KS\ flight length
must be greater than 1\mm, and its direction must form an angle with the \KS\ momentum
vector in this plane that is less than 0.2\rad. A photon is defined as a neutral
cluster of energy greater than 30\mev\ in the EMC 
that agrees in lateral shower shape with an electromagnetic shower.
A \piz\ candidate consists of a pair of photons
with invariant mass in the interval 106 to 153\mevcc.
The \jpsi, \KS\ and \piz\ are constrained to the corresponding nominal masses \cite{ref:pdg}.
\Kstar\ candidates must have $K\pi$ invariant mass within $100\mevcc$
of the nominal $\Kstar (892)$ mass \cite{ref:pdg}.

\B\ mesons are formed from \jpsi\ and \Kstar\ candidates.
For $\B\to\jpsi(K\piz)^*$, \cthetakstar\ is required to be smaller than 0.667.
This reduces
the cross feed (CF) from $\jpsi(K\pipm)^*$ modes, where the \pipm\ is lost, and the self cross feed
(SCF) due to a wrongly reconstructed \piz .
The (S)CF is the most important background source since it tends to peak in the signal
region.

The signal region is defined using two variables. The first is
the difference $\Delta E = E^*_B - E^*_{beam}$
between the candidate
\B\ energy and the beam energy, in the \FourS\ rest frame. The second is the beam-energy
substituted mass
$\mes = (E^2_{exp} - \vec{p}^2_B)^{1/2}$
where, in the laboratory frame, $E_{exp} = \left(s/2 + \vec{p}_B.\vec{p}_i\right)/E_i$ is the
\B\ candidate expected energy, $\vec{p}_B$ its measured momentum, and $(E_i, \vec{p}_i)$ the
$\epem$ initial state four-momentum. $\sqrt{s}$ is the center of mass energy.
For the signal region, $\Delta E$ is required to be between $-70$\mev\ and $+50$\mev\ for
channels involving a \piz, and within $\pm 30$\mev\ otherwise. If several \B\ candidates
are found in an event, the one having the smallest $|\Delta E|$ is retained.
The corresponding \mes\ distributions are shown in Fig.~\ref{fig:bmasses}.

With the signal region defined by $\mes>5.27\gevcc$ and the above $\Delta E$ ranges,
the \B\ reconstruction
efficiencies are 9.9\%, 23.9\%, 17.2\% and 13.8\%
for the \KS\piz , $K^+\pi^-$, \KS$\pi^+$ and $K^+$\piz\ modes, respectively, with
corresponding total yields of
43, 547, 135 and 216 events.
The CF(SCF) contamination levels, obtained from a full simulation of the
\babar\ detector, are 9.9(15.8), 1.2(2.4), 2.4(3.0) and 8.1(15.7)\%
of the pure signal, respectively.

The fit maximizes an unbinned likelihood that 
uses
a probability density function ({\it pdf\/}) that 
depends on angular and \mes\ information.
From the observed \mes\ value, a signal probability is computed with
a Gaussian $G(\mes)$ to describe the signal and a phase-space background function \cite{ref:argusbgk} $F(\mes)$.

The {\it pdf} $\gobs = g(\vec{\omega}_j)\cdot\epsilon(\vec{\omega}_j)/\avEps$
is used to describe signal events;
$\vec{\omega}_j$ represents the angular variables $\cthetatr, \cthetakstar, \phitr$
for event $j$, and $\epsilon(\vec{\omega}_j)$ is the efficiency at $\vec{\omega}_j$.
Rewriting Eq.~\ref{eqn:distrib} as $g = \sum_{i=1}^{6} f_i {\cal A}_i$, where
the ${\cal A}_i (i=1,\ldots,6)$ represent
\azd, \apd, \atd, \pipt, \przp\ and \pizt, the mean efficiency is
$\avEps = \int g\cdot \epsilon\cdot\text{d}\vec{\omega} = \sum_{i=1}^{6}{\cal A}_i\xi_i$,
where the $\xi_i = \int f_i\cdot \epsilon\cdot\text{d}\vec{\omega}$ are constants.
The signal part of the log-likelihood,
$\ln{\cal L}_{signal} = \sum_{j=1}^{N_{obs}} \ln(\gobs(\vec{\omega}_j))$,
where $N_{obs}$ is the number of observed events, becomes
$\ln{\cal L}_{signal} = \sum_{j=1}^{N_{obs}}\ln(g(\vec{\omega}_j)) +\sum_{j=1}^{N_{obs}}\ln(\epsilon(\vec{\omega}_j))
                    -N_{obs}\cdot \ln(\sum_{i=1}^{6}{\cal A}_i\xi_i)$.
Since the $\epsilon(\vec{\omega}_j)$ are constants,
the second term can be discarded.
Only the coefficients $\xi_i$ are required, and detailed representation of the acceptance is
unnecessary \cite{ref:samir}.

The coefficients $\xi_i$ are evaluated with Monte Carlo simulation.
Separate sets
of $\xi_i$ are used for each channel, and for $\ell=e,\mu$. The values of $\xi_i (i=1,2,3)$ 
 are close to that of $\avEps$;
$\xi_1$ is always smallest, especially in channels involving a \piz ,
because of the cut on \cthetakstar .
The values of $\xi_i (i=4,5,6)$,
which are related to the interference terms, are compatible with zero.

The angular dependence of combinatorial background events, $g_{\cal B}^{obs}$,
is described by a {\it pdf\/} similar to that in
Eq.~\ref{eqn:distrib} with amplitudes
$B_{i},i=0,\parallel, \perp,$ and corresponding terms
${\cal B}_i (i=1,\ldots,6)$.

The angular distribution of the (S)CF background is amplitude dependent.
We correct for the effect of this background by
evaluating modified values $\tilde{\xi}_i$ of the $\xi_i$ by including
the (S)CF events, in the \mes\ signal region,  in addition to the signal \cite{ref:samir}.
In contrast to the $\xi_i$, the $\tilde{\xi}_i$ depend on the amplitudes used
in the simulation, but the maximum effect on the fitted amplitudes is found 
to be on the order of $10^{-3}.$
The complete log-likelihood is

\begin{tabular}{lllll}
$\ln{\cal L}$ = & ${\displaystyle \sum_{j=1}^{N_{obs}}}$ & $\ln{\bigg(}$ &  $x\cdot G(\mes_j)\cdot g(\vec{\omega}_j)\;+$      & \\
             &                        &           &  $(1-x)\cdot F(\mes_j) \cdot g_{\cal B}(\vec{\omega}_j)$  & ${\bigg)}$ \\
             & \multicolumn{4}{l}{$- N_{obs} \ln \left( {\displaystyle\sum_{i=1}^{6}}   \tilde{\xi}_i\cdot\left( x \cdot {\cal A}_i +
                                  (1-x)\cdot {\cal B}_i \right) \right) - {\cal N}$} ,
\end{tabular}
where $x$ is the fraction of signal integrated over the \mes\ range 5.2--5.3\gevcc.
The normalization of $g$ and $g_B$ is relaxed in an extended likelihood 
approach \cite{ref:barlow}, with convergence to the required condition
$a^2=\azd+\apd+\atd=1$ imposed through the additional term ${\cal N} =N_{obs} a^2$ while
$|B_{0}|^{2}+|B_{\parallel}|^{2}+|B_{\perp}|^{2}=a^2$ holds by construction.
The fit parameters are the mean and width of $G(\mes)$; the shape parameter of $F(\mes)$;
the fraction $x$; the signal amplitudes and phases
\apd, \azd, \atd, \phit, and \phip; and the corresponding background amplitudes and phases.

\begin{figure}[tbh]
 \begin{center}
    \includegraphics[width=0.93\linewidth]{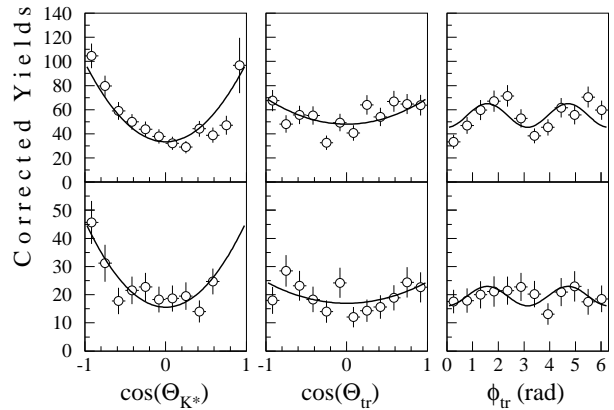}
 \caption{The angular distributions for the channels without (top) and with (bottom) a \piz
for $\mes>5.27\gevcc$.
The data have been background-subtracted and acceptance-corrected. The curves correspond to
 the fit.}
 \label{fig:fitdata}
 \end{center}
\end{figure}

The agreement among the results for the individual decay channels is shown in Table~\ref{tab:reschacha}
while the fit result for the combined samples is
summarized in Table~\ref{tab:results}.
In Fig.~\ref{fig:fitdata}, the result
of the fit to the combined data is compared to the observed angular distributions.
As a check of the fit quality,
Monte Carlo samples were created with the observed angular
distribution and number of events. Subsequent fits gave log-likelihood values in agreement with that obtained
for the data.

\begin{table}[htb]
\begin{center}
\caption{Fitted parameter values for the individual $K\pi$ modes. 
The uncertainties are statistical only.}
\begin{tabular}{ccccc}  \hline\hline
        Quantity        &       \KS\piz         &       $K^+\pi^-$      &       \KS $\pi^+$     &       $K^+$\piz       \\ \hline
        \azd            &       0.65$\pm$0.13   &       0.60$\pm$0.04   &       0.58$\pm$0.07   &       0.55$\pm$0.06   \\
        \atd            &       0.07$\pm$0.11   &       0.17$\pm$0.05   &       0.17$\pm$0.05   &       0.15$\pm$0.08   \\
        \apd            &       0.28$\pm$0.14   &       0.23$\pm$0.05   &       0.25$\pm$0.07   &       0.30$\pm$0.08   \\
        \phit (rad)     &           --          &       $-0.1\pm0.2$    &        0.0$\pm$0.3    &       $-0.4\pm0.4$    \\
        \phip (rad)     &       2.1$\pm$0.7     &        2.5$\pm$0.3    &        2.8$\pm$0.4    &        2.6$\pm$0.5    \\ \hline
\end{tabular}
\label{tab:reschacha}
\end{center}
\end{table}

\begin{table}[htb]
\begin{center}
\caption{Fitted parameter values for the combined data samples.
The first uncertainty is statistical, the second systematic.
Note that $(\phit,\phip)\to(\pi-\phit,-\phip)$ is also a solution.}
\begin{tabular}{cllll}  \hline\hline
        Quantity &      \hspace{2.0cm}          &               &       Value                   \\ \hline
        \azd     &                              &       0.597   & $\pm$ 0.028 & $\pm$ 0.024     \\      
        \atd     &                              &       0.160   & $\pm$ 0.032 & $\pm$ 0.014     \\
        \apd     &                              &       0.243   & $\pm$ 0.034 & $\pm$ 0.017     \\ \hline
        	\phit (rad)    	&		&       $-0.17$ & $\pm$ 0.16  & $\pm$ 0.07      \\
        	\phip (rad)   	&              	&       2.50    & $\pm$ 0.20  & $\pm$ 0.08      \\ \hline
\end{tabular}
\label{tab:results}
\end{center}
\end{table}

\begin{table}[htb]
\begin{center}
\caption{Systematic uncertainties described in the text.}
\begin{tabular}{lccccc}  \hline\hline
                                        &       \azd    &        \atd   &       \apd    &       \phit   &       \phip   \\ \hline
        Simulation stat.                &       0.006   &       0.006   &       0.007   &       0.04    &       0.06    \\
        Backgrounds                     &       0.002   &       0.005   &       0.006   &       0.06    &       0.05    \\
        Tracking and PID\;\;            &       0.002   &       0.006   &       0.004   &       0.00    &       0.02    \\
        $K\pi$ $S$-wave                 &       0.023   &       0.010   &       0.014   &       0.02    &       0.02    \\ \hline
        Total                           &       0.024   &       0.014   &       0.017   &       0.07    &       0.08 \\ \hline
\end{tabular}
\label{tab:systematics}
\end{center}
\end{table}

Systematic uncertainties are detailed
in Table~\ref{tab:systematics}.
Limited simulation statistics (32k events per mode) 
give rise to a systematic uncertainty
in the acceptance and (S)CF corrections (first row).
Monte Carlo simulation has been used to estimate uncertainties due to the assumed form
for the \mes\ and angular distributions of the background (second row). In particular, this
accounts for any possible absorption of the S(CF) background by the $F(\mes)$ function.
Differences between simulated tracking and PID efficiencies 
and measurements obtained with control samples in data
lead to systematic uncertainties (third row) through their impact on acceptance corrections.

\begin{figure}[htb]	
\begin{center}	
 \includegraphics[width=\linewidth]{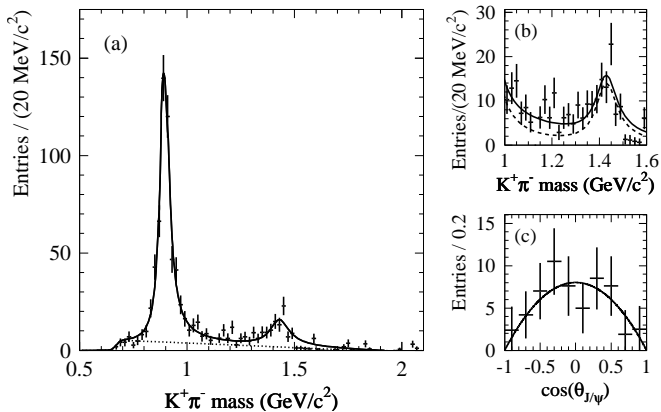}
 \caption{(a) The background-subtracted $K\pi$ mass distribution for the
	  $K^+\pi^-$ channel.
	  The fit is to Breit-Wigner lineshapes having nominal $\Kstar(892)$ and $\Kstar_2(1430)$
	  parameters \cite{ref:pdg} and a second-order polynomial (dotted line). (b) Zoom
	  of the 1--1.6\gevcc\ region of (a); the dashed
	  curve denotes the sum of the Breit-Wigner contributions.
          (c) The background-subtracted \jpsi\ helicity cosine distribution for events with
	  $1.1 < m(K^+\pim) < 1.3$\gevcc;
          the curve represents the fit of a $\sin^2(\theta_\jpsi)$ distribution to the data.
	 }
 \label{fig:heavykstar}
\end{center}
\end{figure}

The $K\pi$ $S$-wave systematic uncertainty (fourth row) is obtained as follows. Although
the $K\pi$ mass distribution for $\B\to\jpsi K\pi$ is dominated by the 
$\Kstar(892)$ (Fig.~\ref{fig:heavykstar}(a)), a significant number of candidates
are at higher mass with
a clear peak at $\sim1.4\gevcc$. The states in this region
that couple strongly to $K\pi$ are the $\Kstar_0(1430)$ and the $\Kstar_2(1430)$ \cite{ref:pdg}.
Since it has width $\sim 300\mevcc$,
the $\Kstar_0(1430)$ alone would yield significantly more events above and
below the peak  than are observed.
The $\Kstar_2(1430)$ alone describes the high mass region but, when combined
with the $\Kstar(892)$ tail, yields too
few events in the 1.1--1.3\gevcc\ range (Fig.~\ref{fig:heavykstar}(b)).
This suggests 
a significant $S$-wave contribution, in which case the recoil
$\jpsi$ has a helicity angle distribution $\sim\sin^2(\theta_\jpsi)$.
The observed behavior (Fig.~\ref{fig:heavykstar}(c)) agrees with this conjecture. This,
together with the absence of $S$-wave above 1.5\gevcc, is
consistent with the mass dependence of the $S$-wave $K\pi$ scattering
amplitude \cite{ref:kpiswave}. If the $K\pi$ $S$-wave in \B\ decay behaves like
this, a coherent $S$-wave amplitude should also be present in the 
$\Kstar(892)$ region; $S$-$P$ interference should occur, which, if ignored, can
affect the $P$-wave amplitudes extracted from the data.

The effect of $S$-wave in the $\Kstar(892)$ region
has been estimated by including a scalar term 
in the total amplitude.
This yields a more complicated angular distribution $g_S$, with ten $f_i$ functions.
A fit of $g_S$  to the data in the
1.1--1.3\gevcc\ region  yields an $S$-wave fraction of
$(62\pm9)\%$, in agreement with the failure of a $P$- and $D$-wave
fit to describe the mass spectrum.
Repeating the analysis using $g_S$, we find the $S$-wave contribution in the  $\Kstar(892)$  region to be
$(1.2\pm0.7)$\%.
The differences in the $P$-wave results with and without $S$-wave are
taken as estimates of systematic uncertainty (Table~\ref{tab:systematics}, fourth row)
since, with the present statistics, the presence of $S$-wave in the
$\Kstar(892)$ region cannot be confirmed.

Table~\ref{tab:comparison} compares our results to those 
of CLEO~\cite{ref:cleo} and CDF~\cite{ref:cdf}. They are consistent,
but the present measurement is significantly more precise.
Longitudinal polarization is seen to dominate and the $P$-wave intensity is small.
If \stwob\ were measured in the $\B\to\jpsi\KS\piz$ channel 
from the decay-time information only, the value of the dilution from the present measurement,
$D_\perp = 0.68\pm 0.07$, would contribute a 10\%
uncertainty.

\begin{table}[htb]
\begin{center}
\caption{Comparison with other experiments. Statistical and systematic uncertainties are added in quadrature.}
\begin{tabular}{lcccc}  \hline\hline
                        &       \azd    &        \atd                   &       \phit           &       \phip      \\ \hline
CLEO \cite{ref:cleo}    &   $0.52\pm 0.08$      &  $0.16\pm 0.09$       &  $-0.11\pm 0.46$      &   $3.00\pm 0.37$ \\
CDF  \cite{ref:cdf}     &   $0.59\pm 0.06$      &$0.13^{+0.13}_{-0.11}$ &  $-0.56\pm 0.54$      &   $2.16\pm 0.47$ \\ \hline
\babar                  &   $0.60\pm 0.04$      &  $0.16\pm 0.03$       &  $-0.17\pm 0.17$      &   $2.50\pm 0.22$ \\ \hline
\end{tabular}
\label{tab:comparison}
\end{center}
\end{table}

Finally, we find that $|\phip|$ differs significantly from $\pi$.
This agrees with the CDF measurement, and indicates a departure from the
factorization of the hadronic currents. 
In addition, there is evidence that
$S$- and $D$-wave amplitude contributions are necessary for a description of the $K\pi$ 
mass spectrum from $\B\to\jpsi K\pi$ decay.


We wish to thank Luis Oliver for many enlightening discussions.
We are grateful for the 
extraordinary contributions of our \pep2\ colleagues in
achieving the excellent luminosity and machine conditions
that have made this work possible.
The collaborating institutions wish to thank 
SLAC for its support and the kind hospitality extended to them. 
This work is supported by
DOE
and NSF (USA),
NSERC (Canada),
IHEP (China),
CEA and
CNRS-IN2P3
(France),
BMBF
(Germany),
INFN (Italy),
NFR (Norway),
MIST (Russia), and
PPARC (United Kingdom). 
Individuals have received support from the Swiss NSF, 
A.~P.~Sloan Foundation, 
Research Corporation,
and Alexander von Humboldt Foundation.



\begin{thebibliography}{99}
\bibitem{foot} Throughout the paper, the use of charge conjugate
modes is implied wherever relevant.
\end{thebibliography}
\end{document}